\documentclass[prl,twocolumn,showpacs,preprintnumbers,superscriptaddress,floatfix,aps]{revtex4-1}

\usepackage{dcolumn}
\usepackage{bbm, bm, epsfig, amsmath, amsfonts, amssymb, wasysym, graphicx, color}
\usepackage[hypertex, bookmarks=false, pdfstartview=FitH]{hyperref}

\newcommand{\ef}{\varepsilon_{\rm F}}
\newcommand{\kf}{k_{\rm F}}

\newcommand{\be}{\begin{equation}}
\newcommand{\ee}{\end{equation}}
\newcommand{\bea}{\begin{eqnarray}}
\newcommand{\eea}{\end{eqnarray}}

\definecolor{darkgreen}{rgb}{0,0.5,0}
\definecolor{purple}{rgb}{0.35,0,0.35}
\definecolor{orange}{rgb}{1,0.5,0}
\definecolor{darkblue}{rgb}{0.1,0.1,.6}
\definecolor{blue}{rgb}{0,0,.8}

\usepackage{soul}

\begin{document}

\relpenalty=9999
\binoppenalty=9999

\title{Dark Continuum in the Spectral Function of the Resonant Fermi Polaron}

\author{Olga Goulko}
\affiliation{Department of Physics, University of Massachusetts, Amherst, MA 01003, USA}
\author{Andrey S. Mishchenko}
\affiliation{RIKEN Center for Emergent Matter Science (CEMS), 2-1 Hirosawa, Wako, Saitama, 351-0198, Japan}
\affiliation{National Research Center ``Kurchatov Institute,'' 123182 Moscow, Russia}
\author{Nikolay Prokof'ev}
\affiliation{Department of Physics, University of Massachusetts, Amherst, MA 01003, USA}
\affiliation{National Research Center ``Kurchatov Institute,'' 123182 Moscow, Russia}
\affiliation{Department of Theoretical Physics, The Royal Institute of Technology, Stockholm SE-10691 Sweden}
\author{Boris Svistunov}
\affiliation{Department of Physics, University of Massachusetts, Amherst, MA 01003, USA}
\affiliation{National Research Center ``Kurchatov Institute,'' 123182 Moscow, Russia}
\affiliation{Wilczek Quantum Center, Zhejiang University of Technology, Hangzhou 310014, China}

\begin{abstract}
We present controlled numerical results for the ground state spectral function of the resonant Fermi polaron in three dimensions. We establish the existence of a ``dark continuum"---a region of anomalously low spectral weight between the narrow polaron peak and the rest of the spectral continuum. The dark continuum develops when the $s$-wave scattering length is of the order of the inverse Fermi wavevector, $a\lesssim 1/\kf$, i.e. in the absence of a small interaction-related parameter when the spectral weight is not expected to feature a near-perfect gap structure after the polaron peak.
\end{abstract}


\maketitle

Ultracold atomic fermions are a versatile and powerful tool to study quantum phenomena in many-body systems. Excellent experimental control and tunability of fermionic mixtures not only provide insight into the physics of such complex systems as electrons in materials or nuclear matter, but also enable a direct realization of fundamental quantum mechanical models. The key observation here is that short-range pairwise interactions near a broad Feshbach resonance \cite{zwerger} are universally characterized by a single dimensionless parameter $\kf a$, where $\kf$ is the Fermi wavevector and $a$ the $s$-wave scattering length, with $\kf a = \infty$ corresponding to the so-called unitary limit.

An important system realized in this way is the resonant Fermi polaron---a spin-down fermion (impurity) in a sea of spin-up fermions \cite{polaronreview, carlosreview} (here we consider three dimensions and equal mass $m$). As a limiting case, it is central for understanding the properties of strongly imbalanced Fermi mixtures. It is also the archetypal example of the dynamic impurity problem featuring strong renormalization of quasiparticle parameters, including changes of fundamental quantum numbers and statistics. At low temperature used in ultracold atom experiments the spin-up subsystem can be regarded as non-interacting, while the impurity gets dressed with particle-hole excitations from the Fermi sea. For sufficiently strong interactions, $\kf a\leq (\kf a)_c=1.11(2)$ \cite{proksvistpolaron1, proksvistpolaron2, polmoltranspunkzwerger, polaronmorachevy, polaroncombescotgiraudleyronas}, a molecular bound state forms between the impurity and one spin-up fermion from the environment.

Most theoretical studies of the Fermi polaron concentrate on computing its ground state energy $E_{\rm p}$, effective mass, and quasiparticle residue $Z$ (modulus square of the overlap between the non-interacting and exact ground state wavefunctions) \cite{proksvistpolaron1, proksvistpolaron2, pilatigiorgini, chevyansatz, chevylobo, bruunmasspolmol, kroiss2d, kroissmassimb, krispolaron, krispolaron2d}. Experiments, on the other hand, probe the spectral function using radiofrequency (rf) and photoemission spectroscopy \cite{expzwierlein, grimmexprepulsive, 2dpolaronskohl, scazza2016}. The quantities of interest are then extracted from the measured spectrum; for instance, $E_{\rm p}$ and $Z$ are given by the frequency and spectral weight of the lowest-frequency sharp peak \cite{expzwierlein}. So far experiments have not yet resolved all the details of the polaron spectral function.

Approximate calculations of the polaron spectral function \cite{polaronfrg, repulsivepolaron, parishlevinsen, kamikado}, which lack control in the absence of a small parameter, report an interval of low spectral weight immediately after the polaron peak, if $\kf a\lesssim 1$. Since, kinematically, there is no restriction on the energy of excited states, the suppressed spectral weight can only be due to anomalously small matrix elements. This is highly surprising given that the quasiparticle residue remains large at $\kf a\lesssim 1$. Only deep in the molecule regime ($\kf a\ll1$) would a near-perfect spectral gap emerge naturally.

In this Rapid Communication, we present controlled diagrammatic Monte Carlo (DiagMC) results for the polaron Green's function, $G_\downarrow$, and its spectral density, $A_\downarrow$, at zero momentum, see Eq.~(\ref{eq:green}). The latter is obtained by the method of Stochastic Optimization with Consistent Constraints (SOCC) \cite{ACpaper}. We take particular care to establish what features of the spectrum can be recovered reliably and what information is inaccessible. Our results show that the interval of anomalously low spectral weight (``dark continuum'') is indeed an unusual physical property of the spectrum.

The simplest effective Hamiltonian of the system reads ($\hbar=1$) \cite{carlosreview}:
\bea
H &=& \sum_{\mathbf{k},\sigma=\uparrow,\downarrow}
\epsilon^{\,}_{\mathbf{k}\sigma}
c^\dagger_{\mathbf{k}\sigma} c^{\,}_{\mathbf{k}\sigma} + \nonumber\\
  &+& \sum_{\mathbf{k},\mathbf{k}',\mathbf{q}}g_0 \, c^\dagger_{\mathbf{k}+\mathbf{q}/2,\uparrow}c^\dagger_{-\mathbf{k}+\mathbf{q}/2,\downarrow}
  c^{\,}_{-\mathbf{k}'+\mathbf{q}/2,\downarrow}c^{\,}_{\mathbf{k}'+\mathbf{q}/2,\uparrow} \;.
\;\;\;\;\;
\eea
We use standard notation for fermionic creation and annihilation operators at momentum $\mathbf{k}$ and spin $\sigma$. The kinetic energy is $\epsilon_{\mathbf{k}\sigma} =k^2/2m-\mu_\sigma$, where $\mu_\uparrow = \ef = \kf^2/2m$ at zero temperature, and $\mu_\downarrow<-\ef$ is an auxiliary control parameter. In what follows we use $\ef$ as the unit of energy. The resonant regime corresponds to the zero-range limit, when the (vanishing) attractive coupling $g_0$ together with the (diverging) ultraviolet cutoff result in a constant value of the $s$-wave scattering length $a$ via an appropriate regularization \cite{revcastin}. Diagrammatically, this limit is reached by replacing bare interaction vertices with resonant T-matrix propagators, which are based on the sum of ladder diagrams. The resulting diagrammatic series is then sampled with DiagMC \cite{proksvistpolaron1, proksvistpolaron2}.

For polaron problems formulated in imaginary time $\tau$ the diagrammatic series converge at any fixed $\tau$ due to the explicit time ordering of interaction vertices, which leads to the $\tau^{2n}/(2n)!$ scaling for the contribution of an individual diagram of order $n$, where $n$ is the number of interaction propagators in a diagram. This scaling overcomes the factorial growth in the number of diagrams. Series convergence is further accelerated by the fact that diagrams have different sign, which leads to strong cancellations of high-order diagrams and reduces their combined contribution \footnote{This might be the reason why the Chevy ansatz \cite{chevyansatz} with a single particle-hole pair already explains many features of the polaron physics with remarkable accuracy \cite{krispolaron}}. Nevertheless, computational complexity does not allow us to sample arbitrarily high orders, and for a controlled solution a systematic extrapolation to the infinite diagram-order limit is necessary. Here we use Abelian resummation that has already been successfully applied to the polaron problem \cite{krispolaron}. The resulting relative error of the polaron Green's function is $\mathcal{O}(10^{-9}-10^{-7})$ at $\tau$ close to zero, $\mathcal{O}(10^{-4}-10^{-3})$ around $\tau=1/\ef$, and a few percent at the largest $\tau$ considered.

The relation between $G_\downarrow(\mathbf{k}, \tau) =-\theta(\tau)\langle c_{\mathbf{k}\downarrow}(\tau)c^\dagger_{\mathbf{k}\downarrow}(0)\rangle$, where $c^{\,}_{\mathbf{k}\downarrow}(\tau)=e^{H\tau}c^{\!}_{\mathbf{k}\downarrow}e^{-H\tau}$ and the average is over the ground state of the system, and $A_\downarrow(\mathbf{k},\omega)$ is standard for linear response theory,
\be
-G_\downarrow(\mathbf{k}, \tau) =\int_0^\infty A_\downarrow(\mathbf{k},\omega)e^{-\omega\tau}d\omega \;.
\label{eq:green}
\ee
The polaron ground state energy, $E_p$, and $Z$-factor control the asymptotic imaginary-time decay of the impurity Green's function at zero momentum,
\be
-G_\downarrow(0,\tau\rightarrow\infty)\rightarrow Ze^{-(E_p-\mu_\downarrow)\tau}\;.
\ee
This implies that the spectral function starts with a sharp peak, $Z\delta(\omega+\mu_\downarrow-E_p)$. Deep in the weak-polaron regime $\kf a\gg(\kf a)_c$ the $Z$-factor approaches unity, meaning that the spectral weight is nearly exhausted by the ground state peak. The particle-hole continuum emerges at higher frequencies and carries the rest of the spectral weight. As there is no kinematic restriction on the allowed energy of excited states, the continuous spectrum starts right after $E_p-\mu_\downarrow$. Its functional form right after the polaron peak is expected to be quadratic in frequency \cite{expzwierlein} (starting at zero spectral density) with a gradual accumulation of the spectral weight towards higher frequencies.
\begin{figure}
\hspace*{-9mm}\includegraphics[width=0.54\columnwidth]{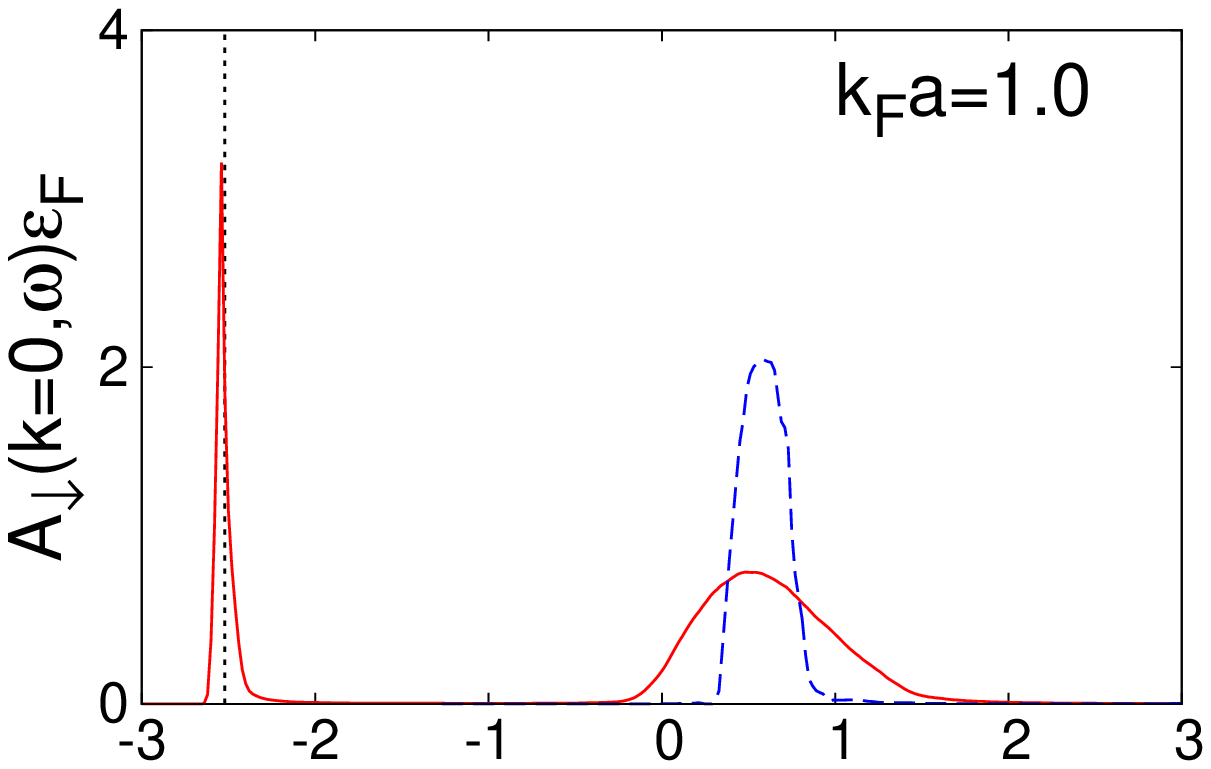}\hspace{-5mm}\includegraphics[width=0.54\columnwidth]{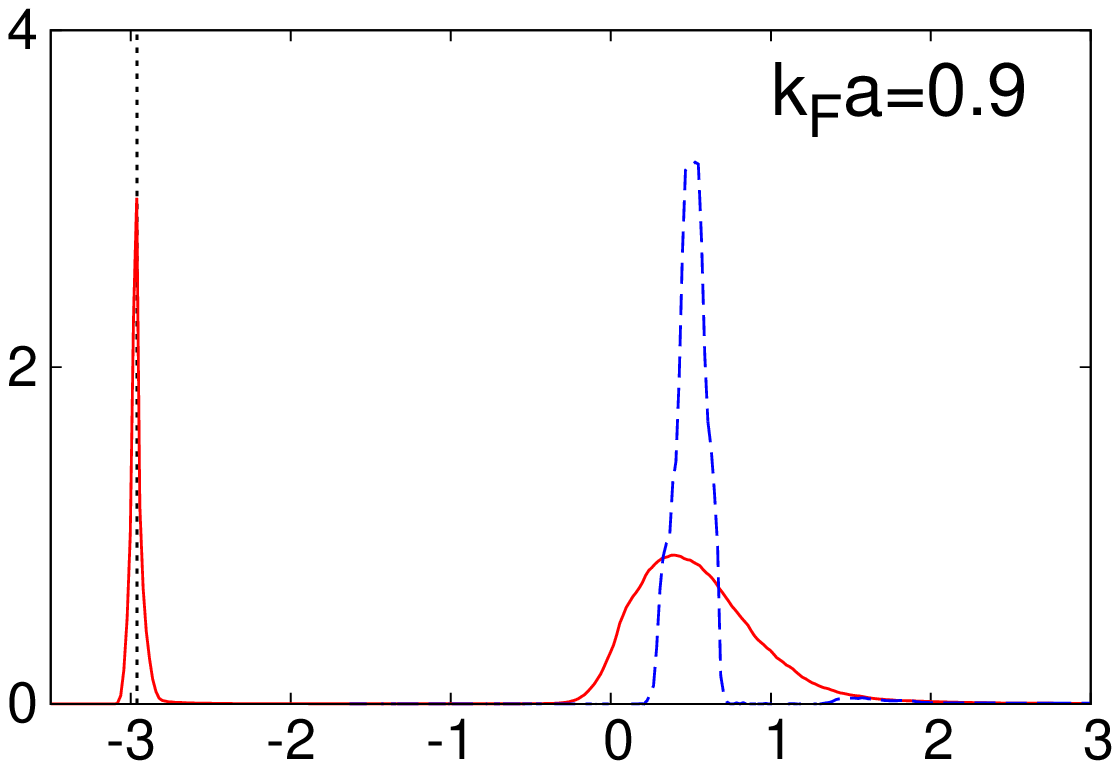}\hspace*{-8mm}\vspace{-3mm}
\hspace*{-9mm}\includegraphics[width=0.54\columnwidth]{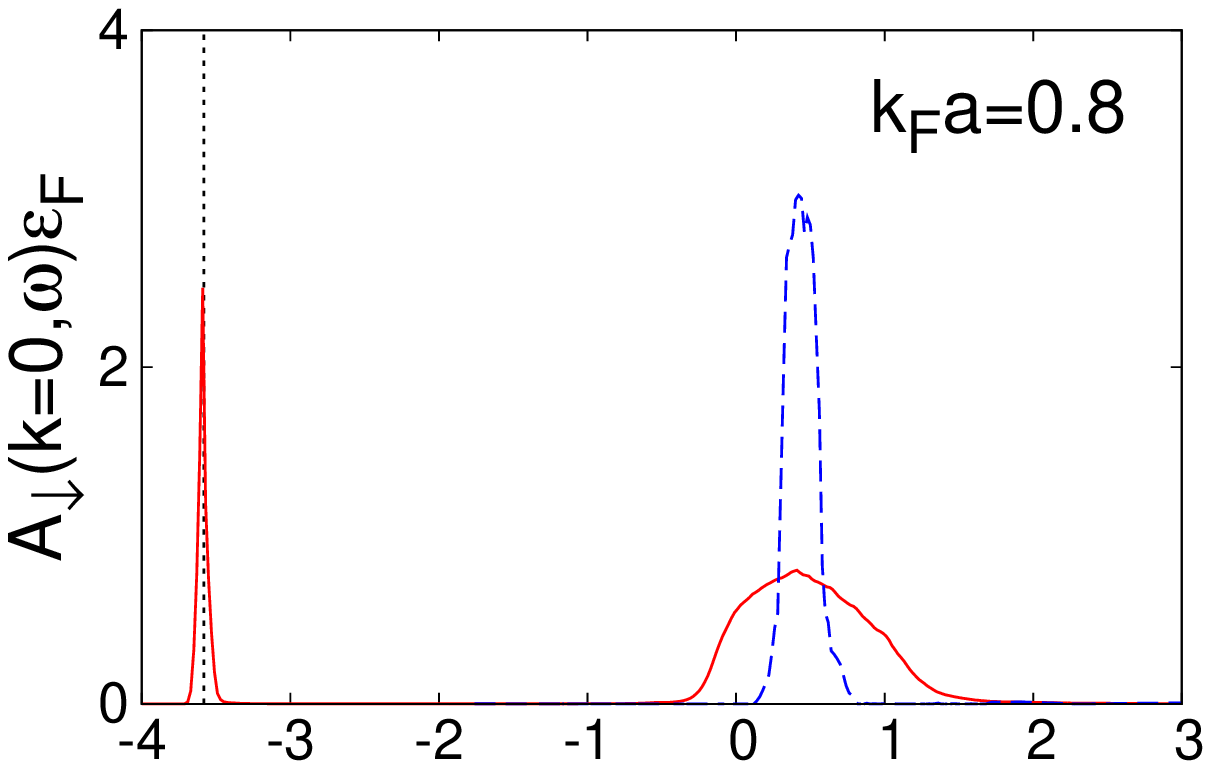}\hspace{-5mm}\includegraphics[width=0.54\columnwidth]{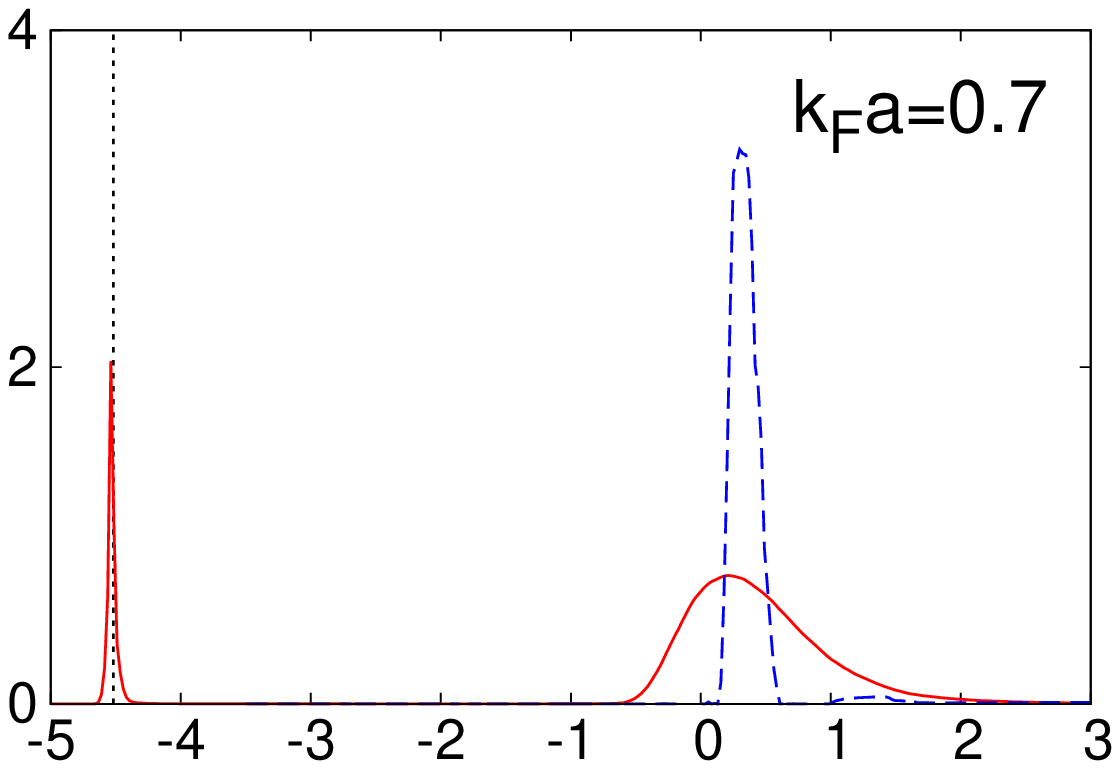}\hspace*{-8mm}\vspace{-3mm}
\hspace*{-9mm}\includegraphics[width=0.54\columnwidth]{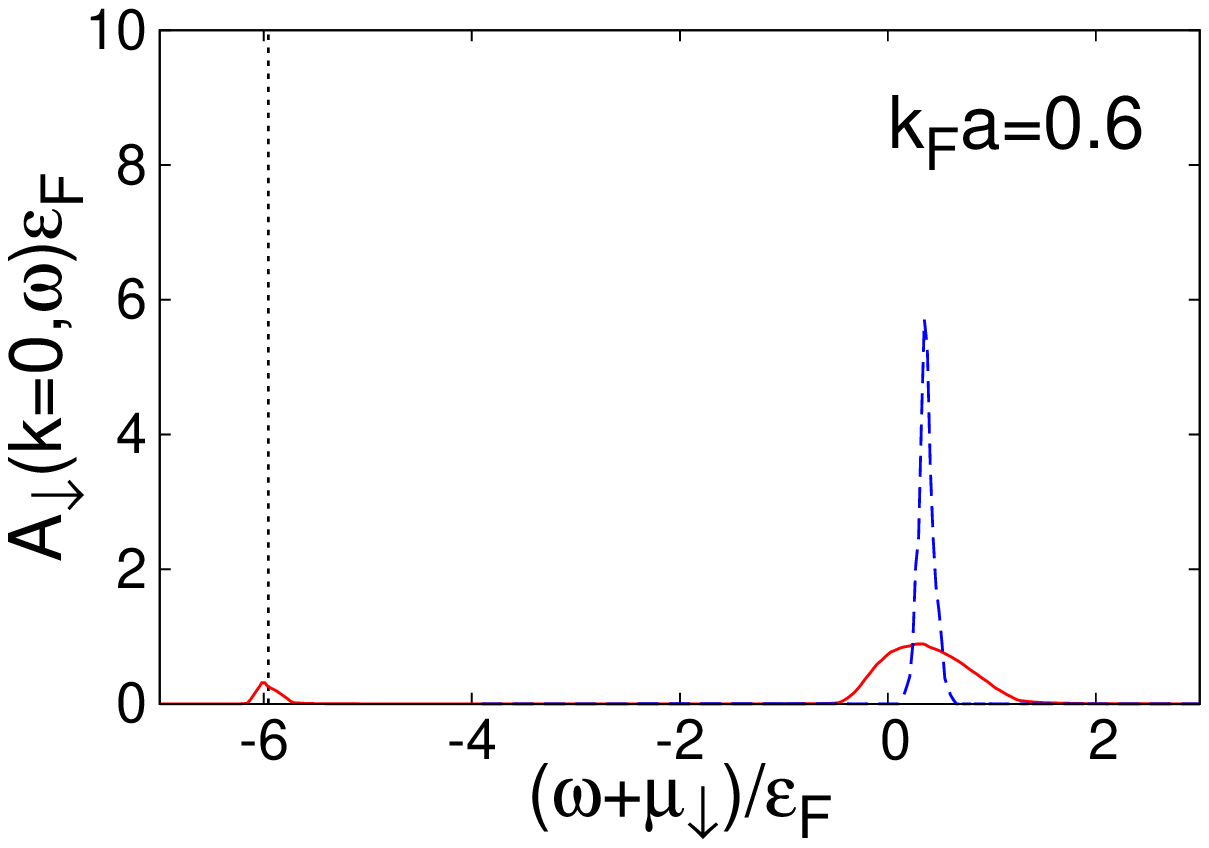}\hspace{-5mm}\includegraphics[width=0.54\columnwidth]{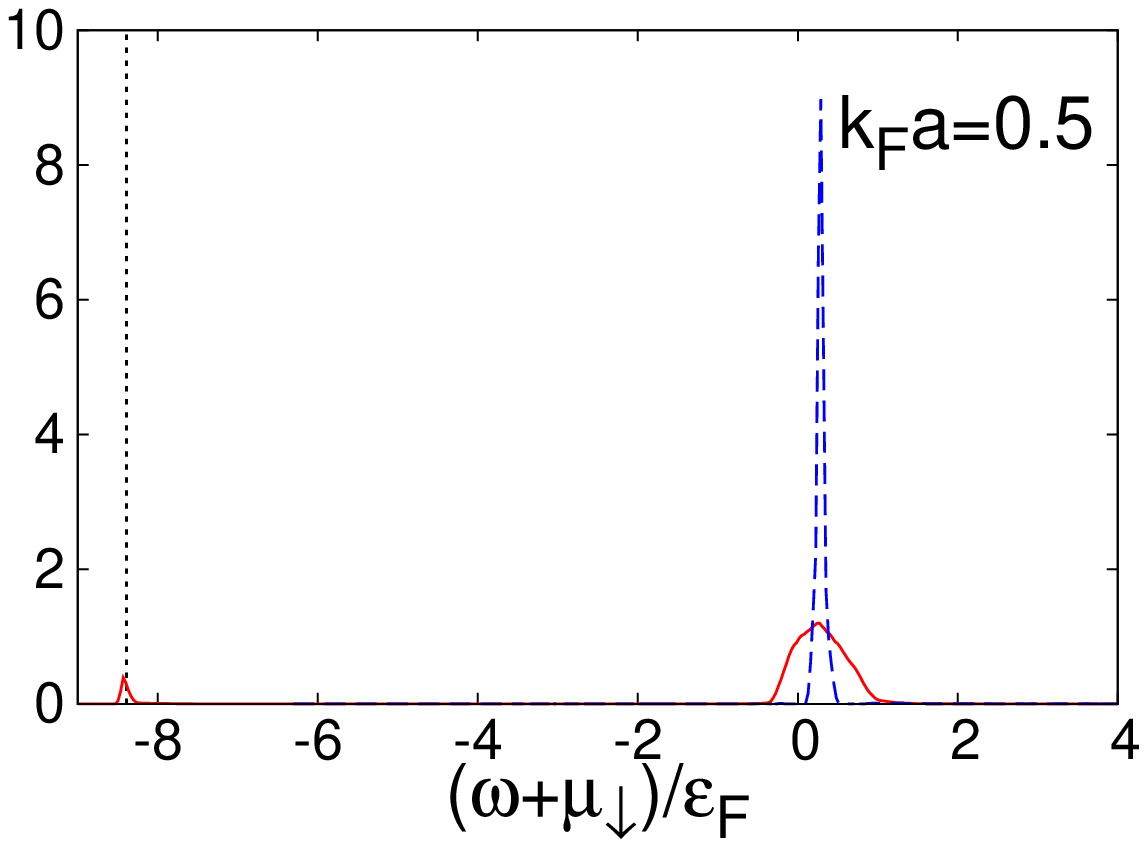}\hspace*{-8mm}
\caption{\label{fig:spectrummol}Spectral function in the molecule regime. The red solid lines show the maximally smooth solution for $A_\downarrow(0,\omega)$ without a $\delta$-function ansatz. The polaron peak is consistent with a $\delta$-function (black short-dashed lines) for all values of $\kf a$ studied. While the smoothest possible solution for the second peak is broad, on the molecule side there exist solutions consistent with a narrow peak, shown by the blue dashed lines.}
\end{figure}
\begin{figure}
\hspace*{-9mm}\includegraphics[width=0.54\columnwidth]{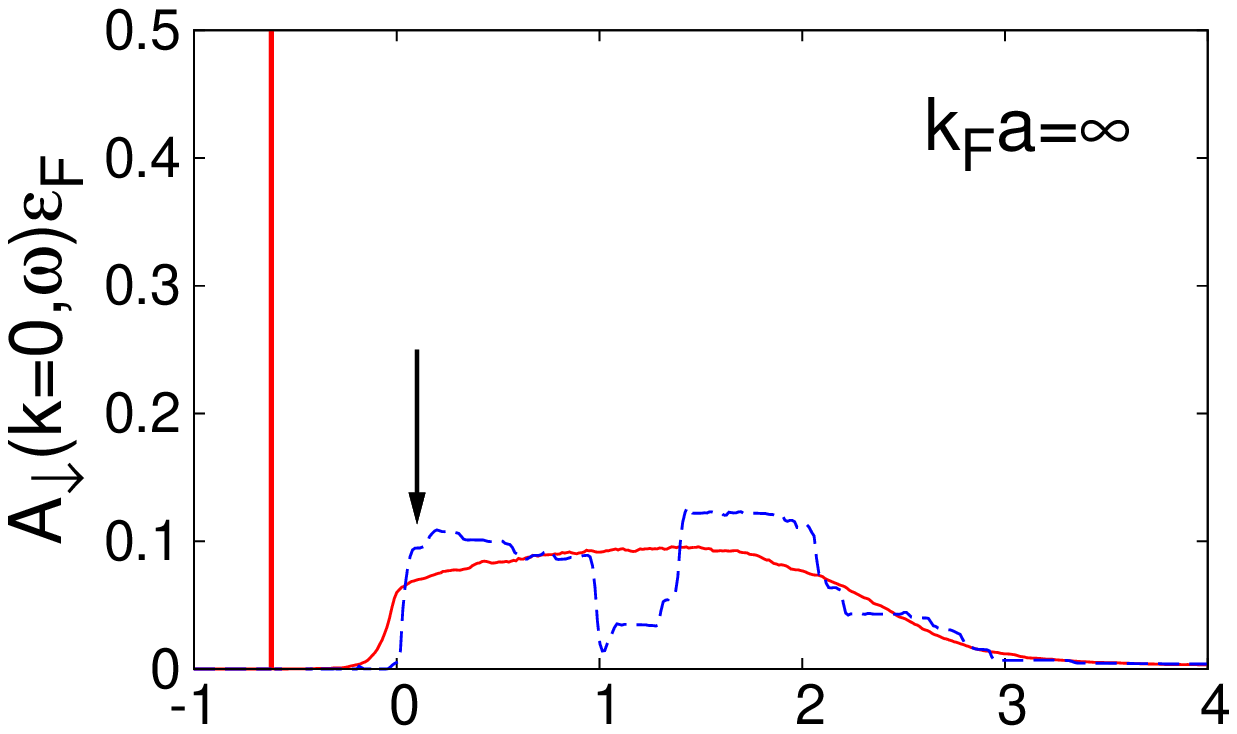}\hspace{-5mm}\includegraphics[width=0.54\columnwidth]{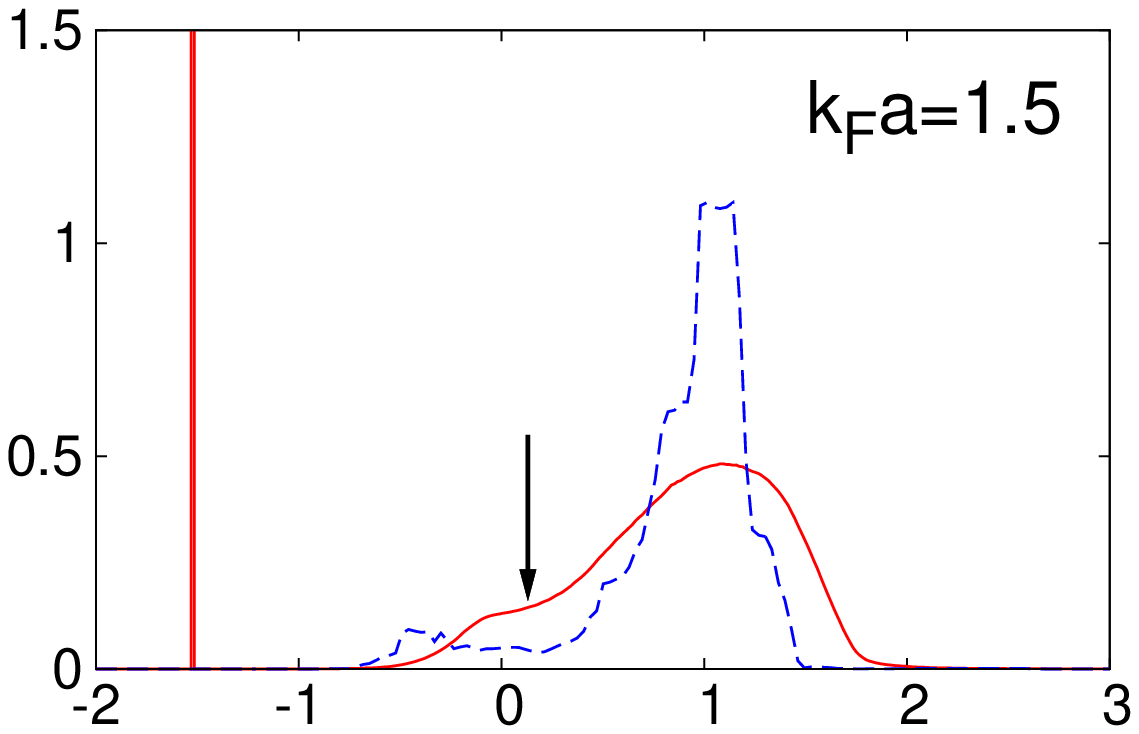}\hspace*{-8mm}\vspace{-5mm}
\hspace*{-9mm}\includegraphics[width=0.54\columnwidth]{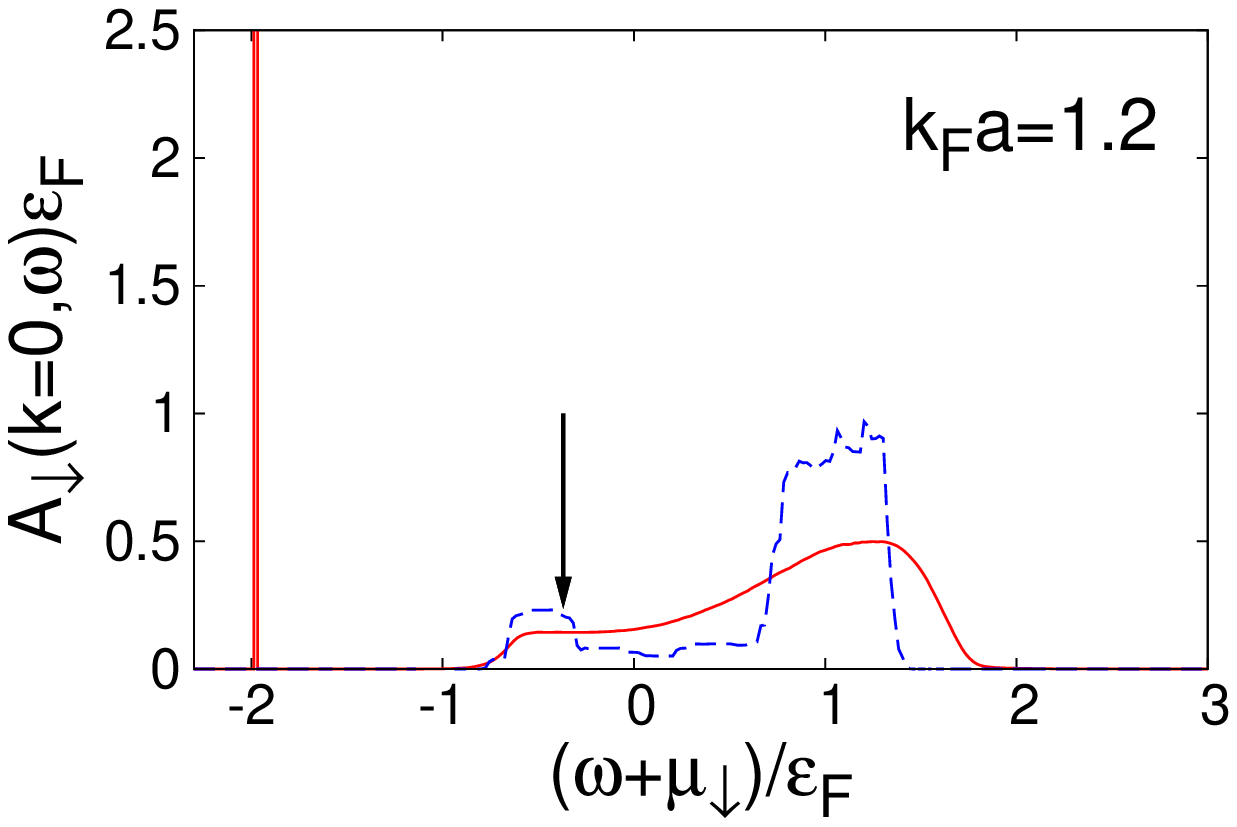}\hspace{-5mm}\includegraphics[width=0.54\columnwidth]{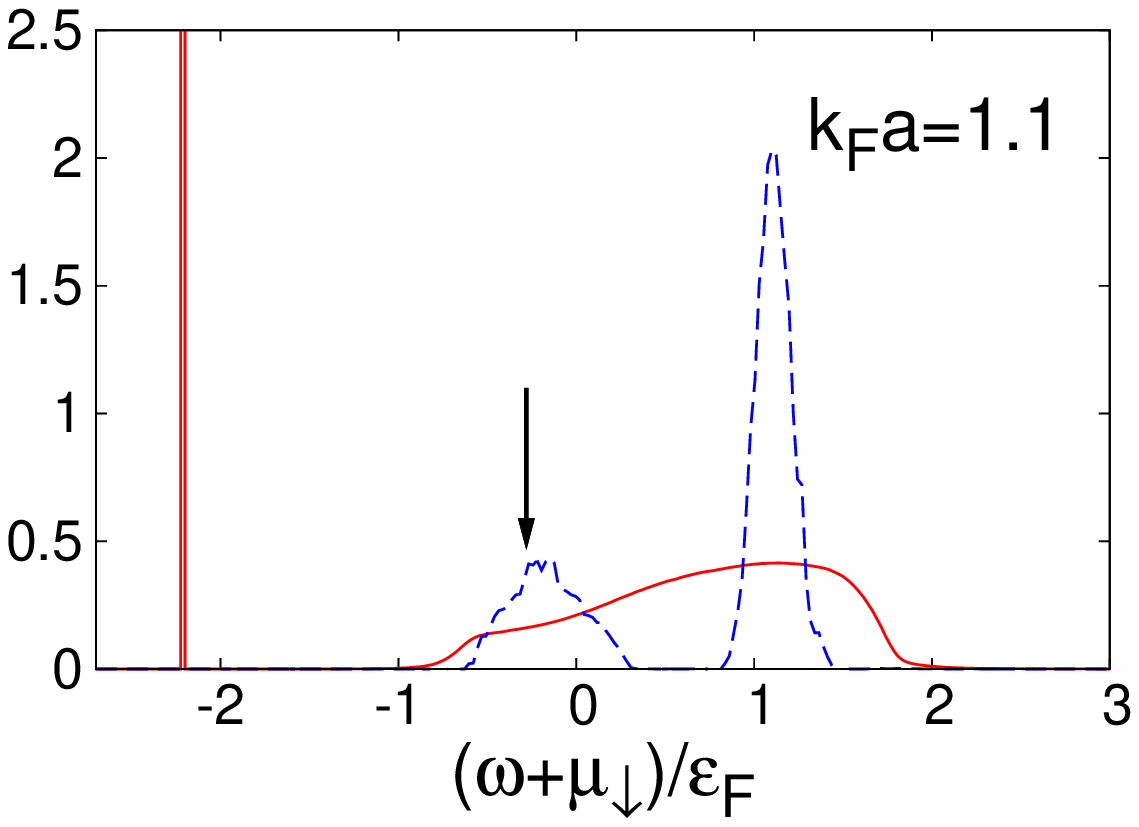}\hspace*{-8mm}
\caption{\label{fig:spectrumpol}Spectral function in the polaron regime, using an explicit $\delta$-function ansatz for the ground state peak. As expected, for $\kf a\gg(\kf a)_c$ the continuous spectrum starts right after the peak, with vanishing spectral density on the approach to the peak, consistent with a vanishing density of states. At sufficiently small $\kf a$, the smoothest possible solution (red solid lines) for the continuous spectrum is a broad peak with a preceding plateau. Within the accuracy of the input data, this structure can be split into two separate peaks (blue dashed lines). The arrow indicates the highest frequency after which the spectrum is incompatible with zero.}
\end{figure}

The SOCC method, described in detail in Ref.~\cite{ACpaper}, produces a smooth solution for the spectrum. But it also allows one to apply additional constraints to explore to what degree the spectrum $A_\downarrow(0,\omega)$ can be altered without compromising the accuracy of the DiagMC data for $G_\downarrow(0, \tau)$. Our philosophy is that the analysis of potential spectral features hidden in the smooth solution is as important as the solution itself. If a feature cannot be changed through additional constraints, the solution is considered robust and its error bars are justified. Otherwise, the feature cannot be reliably resolved.

Figures~\ref{fig:spectrummol} and \ref{fig:spectrumpol} show our results for the spectral function in the molecule and polaron regimes, respectively. For all the studied values of $\kf a$, we observe asymptotic exponential decay of the impurity Green's function with $Z$-factors gradually decreasing as one moves deeper into the molecule regime. This means that the polaron remains a well-defined quasiparticle even well below the transition point. At $(\kf a) \geq (\kf a)_c$, when the polaron is a stable quasiparticle, we use the SOCC protocol with a $\delta$-function ansatz for the polaron peak. At $(\kf a)<(\kf a)_c$, we make no ansatz for the polaron peak, which then acquires a finite width and height determined by the frequency resolution. As $\kf a$ is decreased, the polaron peak moves to lower and lower negative energies, while the bulk of the continuous spectrum only slightly shifts down on the $\omega+\mu_\downarrow \in (0,1)$ interval, and a gap-like region of anomalously low spectral weight develops in between.

\begin{figure}
\includegraphics[width=\columnwidth]{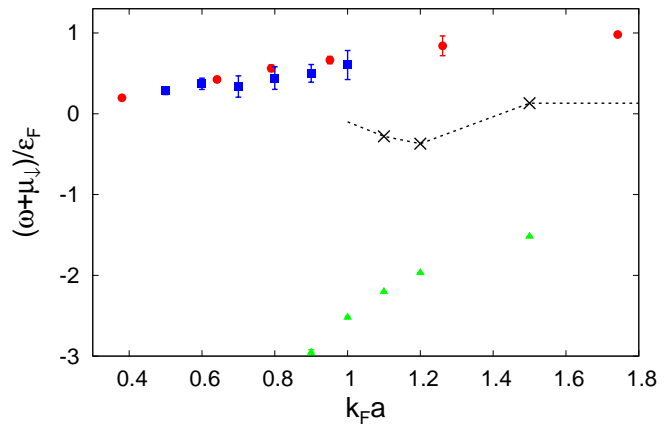}\vspace{-3mm}
\caption{\label{fig:peakenergies}Second peak maximum, $E_2$, of the pulled narrow solution in the molecule regime (blue squares) in comparison with the experiment of Ref.~\cite{scazza2016} (red circles). Black crosses (connected by a dashed line to guide the eye) mark the maximal frequency for the onset of the positive spectrum. Green triangles denote the energy of the polaron peak, $E_p$.}
\end{figure}

The positive energy peak on the molecule side, see Fig.~\ref{fig:spectrummol}, is usually interpreted in the context of the repulsive polaron state \cite{reppolprecursor, polaronfrg, repulsivepolaron, grimmexprepulsive, parishlevinsen, kamikado, scazza2016}. It is extremely desirable to resolve the width of the peak, as the latter can only be associated with a well-defined quasiparticle if the width is much smaller than the characteristic energy scale $\ef$. Unfortunately, the accuracy of our input data is not sufficient to achieve this goal. While the maximally smooth solution indicates a broad peak with width $\sim\ef$, the data for $\kf a<1.1$ are consistent also with a much narrower solution. The latter can be found by applying an additional ``pulling" constraint, which forces the solution to go up (or down) at any specified point, while remaining as structureless as possible otherwise \cite{ACpaper}. Experimental measurements \cite{scazza2016} show that the positive frequency excitation is indeed long-lived in the molecule regime and even on the approach to unitarity. This implies a narrow peak, which was, however, not seen explicitly in the measured spectral response, apparently due to collisional broadening. The measured energies of the second peak agree within error bars with our results for the pulled narrow spectrum, see Fig.~\ref{fig:peakenergies}.

On the polaron side the picture is radically different, see Fig.~\ref{fig:spectrumpol}. On the approach to the transition point, the broad continuous spectrum (the solution obtained under the constraint of maximal smoothness) develops a shoulder preceding the second peak. SOCC analysis reveals that this shoulder could also correspond to a distinct peak. For $(\kf a)\leq 1.5$ the broad second peak can potentially be split into two separate peaks by applying an appropriate pulling constraint. The first of these two peaks is small and, within the limits of our resolution, disappears for $\kf a<1.1$. A potential double-peak structure in the polaron regime was not seen or resolved in the experiment of Ref.~\cite{scazza2016}. With the SOCC protocol, we establish that our input data in the polaron regime are incompatible with a scenario where the shoulder or the small intermediate peak are absent, as opposed to the situation at $\kf a<1.1$. The highest frequency after which the spectrum must be larger than zero is marked by an arrow in the panels of Fig.~\ref{fig:spectrumpol} and plotted in Fig.~\ref{fig:peakenergies}. This frequency is substantially lower than the position of the major second peak, in contrast to what we see on the molecule side. Deep in the polaron regime, see the top left panel of Fig.~\ref{fig:spectrumpol}, the spectrum can no longer be split into two clearly distinct peaks. The spectral weight starts accumulating right after the polaron peak, as expected.

The region of low spectral weight between the two peaks has been seen in variational calculations \cite{repulsivepolaron, parishlevinsen} and functional Renormalization Group studies \cite{polaronfrg, kamikado}. Our calculations confirm that it is indeed a physical effect in the three-dimensional broad-resonance equal-mass case, not an artifact of uncontrolled approximations. The integrated spectral weight of this region remains anomalously low even when a pulling constraint is applied in the SOCC protocol attempting to maximally increase it. This structure can thus be well resolved from the input data and the error bars are controlled.

The ``dark" spectral continuum refers to few-body states (containing a polaron or molecule plus a number of particle-hole pairs) that are expected to get excited with substantial probability when the impurity particle is created, but instead remain invisible in the spectral function. In the absence of a small (or large) dimensionless parameter at $\kf a\sim1$, an \emph{a priori} order-of-magnitude estimate of the overall spectral weight associated with this few-body continuum is that of the polaron $Z$-factor. With the gradual decrease of the latter, the dark continuum states are not expected to have their integrated spectral weight dramatically smaller than that of the stable or metastable polaron, provided $\kf a$ is not small. 

\begin{figure}
\includegraphics[width=\columnwidth]{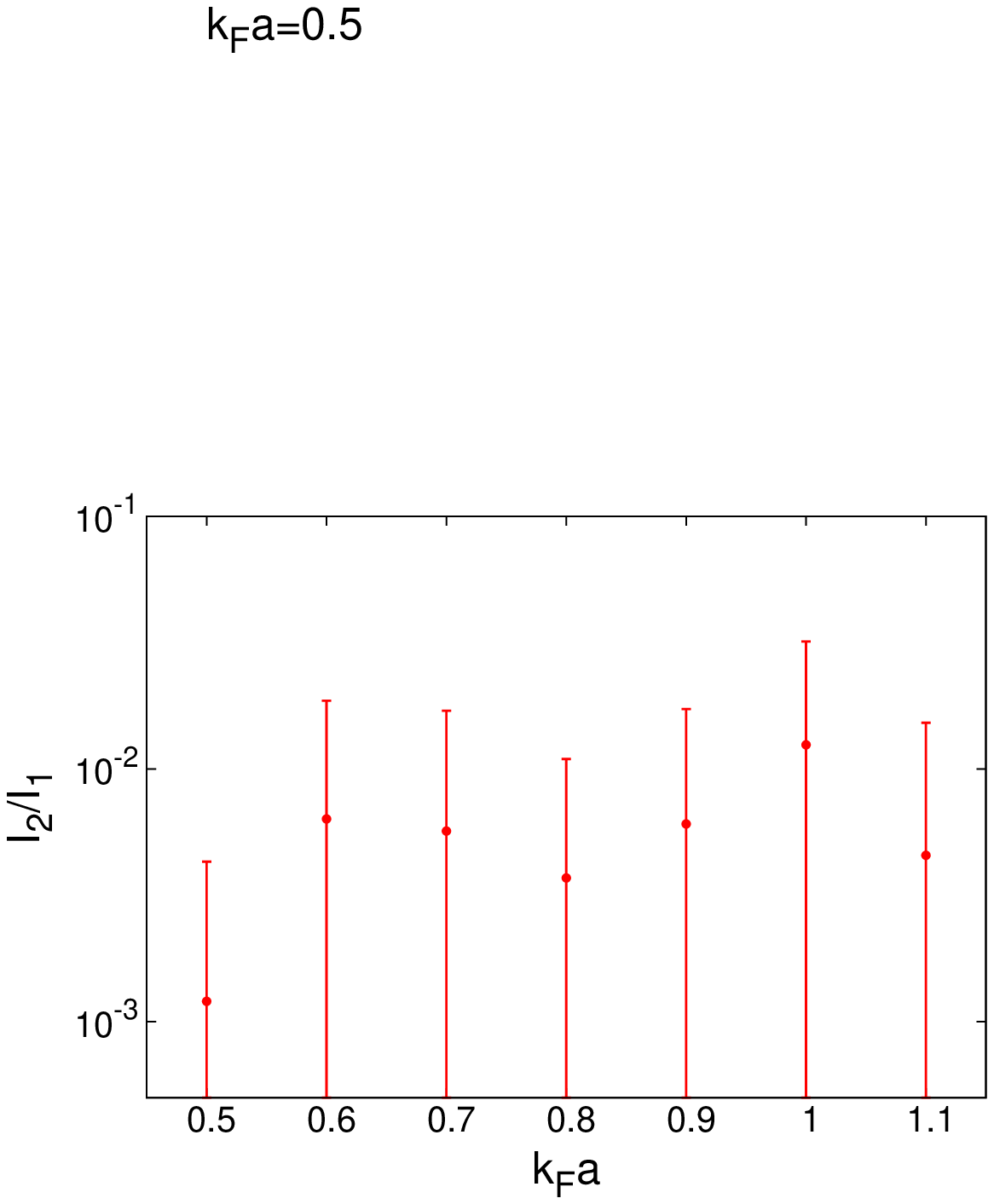}\vspace{-3mm}
\caption{\label{fig:darkcont} Integrated spectral weight of the dark continuum, $I_2$, in units of integrated spectral weight of the polaron peak, $I_1$.}
\end{figure}

To quantify the dark continuum, we establish controlled estimates on its integrated spectral weight, see Fig.~\ref{fig:darkcont}. We start by determining the positions of the two maxima in the spectral function, $E_p$ and $E_2$, and the interval $\Delta E=E_2-E_p$. Without a $\delta$-function ansatz, the polaron peak has a finite width due to the frequency-resolution. We characterize its spectral weight as $I_1=\int_{0}^{E_p-\mu_\downarrow+\Delta E/4} A_\downarrow(0,\omega)d\omega$, which is very close to the value of the $Z$-factor. The dark continuum region is then defined to extend from $E_p-\mu_\downarrow+\Delta E/4$ to the midpoint between the peaks, $E_p-\mu_\downarrow+\Delta E/2$. Our choice of intervals is rather arbitrary, but illustrates well the orders of magnitude difference between the spectral integrals of the polaron peak, $I_1\approx Z$, and the dark continuum, $I_2=\int_{E_p-\mu_\downarrow+\Delta E/4}^{E_p-\mu_\downarrow+\Delta E/2} A_\downarrow(0,\omega)d\omega$.

In conclusion, we have presented controlled first-principles evidence for an unusual shape of the spectral function of the resonant Fermi polaron. Based on the kinematics of the problem and the absence of small parameters, one would expect to see the following features on the molecular side $\kf a < (\kf a)_c$: 
(i) a significant broadening of the polaron peak due to the decay into the molecule state for $\kf a\ll(\kf a)_c$; (ii) the spectral continuum starting directly after the (broadened) polaron peak; (iii) progressive loss of the spectral weight of the polaron peak to the surrounding molecule-hole continuum as $\kf a$ decreases. 
Instead, essentially the opposite happens: the polaron peak stays sharp and contrasts even stronger with the suppressed background. In particular, this means that the spectral function alone is insufficient to determine the molecule transition point.

The accuracy of our numerical data is sufficient to establish the existence of the dark continuum for $\kf a < (\kf a)_c$. What we cannot reliably resolve is the width and the structure of the higher-frequency part of the spectral function. A substantial increase in the resolution of the numerical spectral analysis would require a reduction of the error bars on the input data by at least an order of magnitude (implying a several orders of magnitude increase of computation time), which is not feasible at present. Experimental measurements of the decay rate reveal a metastable repulsive polaron state with a long lifetime even near unitarity \cite{scazza2016}. This suggests a very simple scenario when the spectral density is essentially exhausted  by two narrow quasiparticle peaks. The results of Fig.~\ref{fig:spectrummol} show that this picture is consistent with our data at $k_Fa\leq1$. However, the results of Fig.~\ref{fig:spectrumpol} clearly indicate that it fails at $k_Fa \geq 1.1$. Here, the higher-frequency part of the spectral function cannot be reduced to a single narrow peak. Even if a peak is present, there should be additional contributions to the spectral density.

We thank T.~Enss, P.~Kroiss, C.~Lobo, L.~Pollet, A.~Recati, G.~Roati, F.~Scazza, R.~Schmidt, K.~Van~Houcke, F.~Werner and M.~Zaccanti for helpful discussions and F.~Scazza and M.~Zaccanti also for providing us with their experimental data for comparison. This work was funded by the NSF under Grant No. PHY-1314735, by the ImPACT Program of the Council for Science, Technology and Innovation (Cabinet Office, Government of Japan), and Stiftelsen Olle Engkvist Byggm\"{a}stare Foundation.

\bibliography{allbib,mylocalbib}

\end{document}